\documentclass[useAMS,usenatbib]{mn2e}
\newcommand{\be}{\begin{equation}}
\newcommand{\ee}{\end{equation}}
\newcommand{\ba}{\begin{eqnarray}}
\newcommand{\ea}{\end{eqnarray}}

\title{Gamma-ray burst internal shocks with magnetization}
\author[]{Y. Z. Fan$^{1,2 \star}$, D. M. Wei$^{1,2}$ and Bing
Zhang$^{3,4}$
\thanks{E-mail: yzfan@pmo.ac.cn(YZF); dmwei@pmo.ac.cn(DMW);
bzhang@physics.unlv.edu(BZ)} \\
$^1${\sl Purple Mountain Observatory, Chinese Academy of
Science, Nanjing 210008, China}\\
       $^2${\sl National Astronomical
Observatories, Chinese Academy of Sciences, Beijing 100012,
China}\\
$^3${\sl Department of Astronomy and Astrophysics, Pennsylvania
State
University, University Park, PA 16082, USA}\\
$^4${\sl Department of Physics, University of Nevada, Las Vegas,
NV 89154, USA}}

\begin{document}

\maketitle
\begin{abstract}

We investigate Gamma-ray Burst (GRB) internal shocks with moderate
magnetization, with the magnetization parameter $\sigma$ ranging
from 0.001 to 10. Possible magnetic dissipation in the stripped
magnetized shells is also taken into account through introducing a
parameter $k$ ($0<k\leq 1$), which is the ratio of the electric
field strength of the downstream and the upstream.  By solving the
general MHD jump conditions, we show that the dynamic evolution of
the shock with magnetic dissipation is different from the familiar
one obtained in the ideal MHD limit. As long as the relative
velocity between the two magnetized shells is larger than the
corresponding Alfven velocities in both shells, strong internal
shocks still exist for $\sigma\gg 1$, which can effectively tap
kinetic energy into radiation. However, in the ideal MHD limit
($k=1$), the upstream magnetic energy can not be converted into
the downstream thermal energy so that the GRB radiation efficiency
is low. This is likely inconsistent with the current GRB data.
With magnetic dissipation, e.g., $k\leq 0.5$ the range of $k$ is
constrained given a particular upstream-downstream Lorentz factor
$\gamma_{21}$ and a magnetization parameter $\sigma$), a
significant fraction of the upstream magnetic energy can be
converted into the prompt $\gamma-$ray emission. At the typical
internal shock radius, the characteristic synchrotron emission
frequency in the magnetic dissipation dominated case is however
too large ($\propto \sigma^2$) compared with the data if
$\sigma\gg 1$.  On the other hand, as long as the ordered magnetic
field is stronger than or at least comparable with the random one
generated in the internal shocks, a net linear polarization $\geq
30\%$ results. In view of the possible high degree of linear
polarization of GRB 021206 and the identification of a possible
highly magnetized flow in GRB 990123 and GRB 021211, we suggest
that a mildly magnetized internal shock model ($0.01<\sigma<1$)
with moderate magnetic dissipation is a good candidate to explain
the GRB prompt emission data.

\end{abstract}


\section{Introduction}
Tremendous advances to understand the Gamma-ray bursts (GRBs), one
of the greatest enigmas in high energy astrophysics have been
achieved in the past seven years (see M\'{e}sz\'{a}ros 2002; Cheng
\& Lu 2001; Zhang \& M\'{e}sz\'{a}ros 2004 for reviews). However,
the nature of the GRB central engine is still unclear. In the
conventional fireball model, a GRB is powered by the collisions of
non-magnetized shells with variable Lorentz factors, i.e., the
internal shocks (Paczy\'{n}ski \& Xu 1994; Rees \&
M\'{e}sz\'{a}ros 1994; Kobayashi, Piran \& Sari 1997; Pilla \&
Loeb 1998; Daigne \& Mochkovitch 1998; Guetta, Spada \& Waxman
2001). The magnetic field involved in the synchrotron radiation
model for the prompt $\gamma-$ray emission is usually interpreted
as being generated in the internal shocks, and is randomly
oriented in the shock plane with small coherence scale, so that
there is no net polarization expected in the prompt $\gamma-$ray
emission (Medeveder \& Loeb 1999). Recently two pieces of
independent evidence suggest that the GRB central engine might be
strongly magnetized. First, the detection of the very high linear
polarization of GRB 021206 (Coburn \& Boggs 2003, but see Rutledge
\& Fox 2004) suggests that the magnetic field involved in the
synchrotron radiation could be globally ordered (e.g. Coburn \&
Boggs 2003; Lyutikov, Pariev \& Blandford 2003; Granot 2003;
Granot \& K\"{o}nigl 2003), although some alternative explanations
such as the Compton scattering model (e.g., Shaviv \& Dar 1995;
Lazzati et al. 2000; Eichler \& Levinson 2003) and the narrow jet
model (Waxman 2003) remain. Second, analysis of the two
well-studied optical flashes from GRB 990123 and GRB 021211 reveal
that the magnetic fields in the reverse shock region are stronger
than that in the forward shock region, so that the GRB outflows
are likely magnetized (Fan et al. 2002; Zhang, Kobayashi \&
M\'{e}sz\'{a}ros 2003; Kumar \& Panaitescu 2003). The current
models involving ordered magnetic fields for GRBs invoke a
Poynting flux dominated outflow (e.g. Usov 1994; Thompson 1994;
M\'{e}sz\'{a}ros \& Rees 1997; Spruit, Daigne \& Drenkhahn 2001),
in which the ratio of the electromagnetic energy flux to the
kinetic energy flux of the baryons (i.e., $\sigma$) is of order
100 or more, and the GRB prompt emission is envisaged to be due to
some less familiar magnetic dissipation process. In reality, a GRB
event likely involves a ``hot component'' as invoked in the
traditional fireball model (e.g. due to neutrino annihilations),
whose interplay with the ``cold'' Poynting flux component would
allow the $\sigma$ value to vary in a wide range (e.g., Rees \&
M\'{e}sz\'{a}ros 1994; Zhang \& M\'{e}sz\'{a}ros 2002).  On the
other hand, numerical simulations and statistic analysis of GRBs
suggest that the internal shock model is preferred (e.g.,
Kobayashi et al. 1997; Lloyd-Ronning, Petrosian \& Mallozzi 2000;
Guetta et al. 2001; Amati et al. 2002; Zhang \& M\'{e}sz\'{a}ros
2002; Wei \& Gao 2003).  Motivated by these facts, in this paper,
we investigate the GRB internal shocks with magnetization. Spruit
et al. (2001) have discussed this topic briefly by taking the
energy and momentum conservation of two magnetized shell
collision. More detailed treatments are needed.

This paper is structured as follows. We first discuss the MHD
90$^{\rm o}$ shock jump conditions with magnetic dissipation, both
analytically ($\S{2.1}$) and numerically ($\S{2.2}$).
We then (\S3) discuss the fast ejecta --
slow ejecta interaction, in particular for an ejecta with moderate
magnetization (i.e., $\sigma<10$), and calculate the prompt
synchrotron emission from such moderately magnetized internal
shocks. Our results are summarized in \S4.

\section{Investigation of the relativistic MHD $90^{\rm 0}$
shocks with magnetic dissipation}

It is generally believed that a GRB involves a rapidly rotating
central engine. If the magnetic fields from the engine are frozen
in the expanding shell, the radial magnetic field component
decreases more rapidly with radius ($\propto r^{-2}$) than the
toroidal field component ($\propto r^{-1}$). At the internal shock
radius, the frozen-in field is likely dominated by the toroidal
component, so that the field lines are essentially perpendicular
to the shock normal direction, i.e. one has a $90^{\rm o}$ shock.

\subsection{The general Jump Conditions}
A rigorous analytical treatment of the MHD $90^0$ shock jump
conditions has been presented by Zhang \& Kobayashi (2004,
hereafter ZK04) recently. Similar to ZK04, here we present a
rigorous analytical solution for the $90^0$ shock jump condition
with magnetic dissipation. The main difference between this work
and ZK04 is the following. In their ideal MHD (i.e., without
magnetic dissipation) limit,
the electric field in the shock frame is continuous across the
shock. This is no longer true in the presence of magnetic energy
dissipation, which may result, e.g., from magnetic reconnection in
the shock front (e.g. Levinson \& van Putten 1997; Lyubarsky
2003). In this work, following the treatment of Lyubarsky (2003)
on the termination shock in a stripped pulsar wind, we introduce a
parameter
\begin{equation}
 k\equiv \beta_{\rm 2s}B_{\rm 2s}/\beta_{\rm 1s}B_{\rm 1s},~~~0\leq k \leq 1
\end{equation}
to describe the potentially important but poorly understood
magnetic dissipation process, where $\beta_{\rm 1s}$ and
$\beta_{\rm 2s}$ ($B_{\rm 1s}$ and $B_{\rm 2s}$) are the
velocities (magnetic field strength) of the upstream and
downstream regions measured in the shock frame, respectively. In
principle, $k$ is constrained by the stripped part of the Poynting
flux, which is assumed as a free parameter ranging from 0 to 1 in
the current work.

Now, following ZK04, we consider a relativistic shock that
propagates into a magnetized ejecta. In the following analysis,
the unshocked region (upstream) is denoted as the region 1, the
shocked region (downstream) is denoted as the region 2, and the
shock itself is denoted as ``$s$''\footnote{Notice that such a
notation system is only valid for \S{2} and the Appendix. When
discussing the GRB problem, i.e. the fast shell-slow shell
interaction (\S{3}), we introduce different meanings for the
subscript numbers.}. Hereafter $Q_{\rm ij}$ denotes the value of
the quantity $Q$ in the region ``$i$'' in the rest frame of
``$j$'', and $Q_{\rm i}$ denotes the value of the quantity $Q$ in
the region ``$i$'' in its own rest frame. For example,
$\gamma_{21}$ is the Lorentz factor of regions 2 relative to
region 1, $\beta_{\rm 1s}$ is the velocity (in unit of the speed
of light $c$) of region 1 relative to the shock front, $B_{\rm
2s}$ ($E_{\rm 2s}$) is the magnetic (electric) field strength of
the region 2 measured in the rest frame of the shock, while $B_1$
is the comoving magnetic field strength in the region 1, etc. In
the presence of the magnetic energy dissipation, the familiar
relativistic 90$^{\rm o}$ shock Rankine-Hugoniot relations
(Hoffmann, De \& Teller 1950; Kennel \& Coroniti 1984, hereafter
KC84) take the general form
\begin{eqnarray}
&n_1 u_{\rm 1s}         =  n_2 u_{\rm 2s}~, \label{jump1}\\
&E_{\rm 1s}= \beta_{\rm 1s} B_{\rm 1s} ; E_{\rm 2s}= \beta_{\rm 2s} B_{\rm 2s}~, \label{EB} \\
&\gamma_{\rm 1s} \mu_1 + \frac{ E_{\rm 1s} B_{\rm 1s}}{4\pi n_1
u_{\rm 1s}}=\gamma_{\rm 2s} \mu_2 + \frac{ E_{\rm 2s} B_{\rm 2s}}{4\pi n_2 u_{\rm 2s}}~, \label{jump2}\\
&\mu_1 u_{\rm 1s}+\frac{p_1}{n_1 u_{\rm 1s}} + \frac{B_{\rm
1s}^2+E_{\rm 1s}^2}{8\pi n_1u_{\rm 1s}}
   =  \mu_2 u_{\rm 2s}+\frac{p_2}{n_2 u_{\rm 2s}}
  + \frac{B_{\rm 2s}^2+E_{\rm 2s}^2}{8\pi n_2 u_{\rm 2s}}~, \label{jump3}
\end{eqnarray}
where $\beta$ denotes the dimensionless velocity,
$\gamma=(1-\beta^2)^{-1/2}$ denotes the Lorentz factor, and
$u=\beta \gamma$ denotes the radial four velocity. Hereafter, $n$,
$e$, $p=(\hat\Gamma-1) e$ denote the number density, internal
energy and thermal pressure, respectively, and $\hat\Gamma$ is the
adiabatic index. The enthalpy is $n m_{\rm p} c^2+e+p$, and the
specific enthalpy can be written as \be \mu = m_{\rm p} c^2 +
\frac{\hat\Gamma}{\hat\Gamma-1}\left(\frac{p}{n}
\right),\label{mu} \ee where $m_{\rm p}$ is the proton mass and
$c$ is the speed of light. It is convenient to define a parameter
\be \sigma_{\rm i}=\frac{B_{\rm i}^2}{4\pi n_{\rm i}\mu_{\rm
i}}=\frac{B_{\rm is}^2}{4\pi n_{\rm i}\mu_{\rm i} \gamma_{\rm
is}^2}, \ee to denote the degree of magnetization in each region.
The magnetization parameter in the upstream region ($\sigma_1$) is
a more fundamental parameter, since it characterizes the
magnetization of the flow itself. We therefore define
\begin{equation}
\sigma \equiv \sigma_1 = \frac{B_{\rm 1s}^2}{4\pi
n_1\mu_1\gamma_{\rm 1s}^2}. \label{sigma}
\end{equation}
In our problem, we are interested in a ``cold'' upstream flow,
i.e., $e_1=p_1=0$, so that $\mu_1=m_{\rm p} c^2$. This is the only
assumption made in the derivation. With equation (\ref{jump2}),
the thermal Lorentz factor of the downstream particles is
\begin{equation}
\frac{e_2}{n_2 m_{\rm p} c^2} = {1\over \hat{\Gamma}}
\{{\gamma_{\rm 1s}\over \gamma_{\rm 2s}}[1+\sigma(1-k^2{\beta_{\rm
1s}\over \beta_{\rm 2s}})]-1\}, \label{e/n}
\end{equation}
where $\gamma_{\rm 2s}(\gamma_{21},\sigma,k)$ is a function of
$\gamma_{21}$, $\sigma$, and $k$, and can be solved once
$\gamma_{21}$, $\sigma$ and $k$ are known. The equation governing
$\gamma_{\rm 2s}(\gamma_{21},\sigma,k)$ reads (see Appendix for
derivation, see also Lyubarsky 2003)
\begin{eqnarray}
&&\gamma_{\rm 1s}u_{\rm 1s}(1+{1\over \sigma})[\beta_{\rm
1s}-\beta_{\rm 2s}-{(\hat{\Gamma}-1)\over \hat{\Gamma}}{1\over
u_{\rm 2s}\gamma_{\rm 2s}}]+{1\over
2}\nonumber\\
&&=k^2{(2-\hat{\Gamma})\over 2\hat{\Gamma}}{u_{\rm 1s}^2\over
u_{\rm 2s}^2}-{(\hat{\Gamma}-1)\over \hat{\Gamma}\sigma}{u_{\rm
1s}\over u_{\rm 2s}},\label{Num1}
\end{eqnarray}
where $\beta_{\rm 1s}=(\beta_{\rm
2s}+\beta_{21})/(1+\beta_{21}\beta_{\rm 2s})$, $\gamma_{\rm
1s}=\gamma_{\rm 2s}\gamma_{21}(1+\beta_{21}\beta_{\rm 2s})$ and
$u_{\rm 1s}=\beta_{\rm 1s}\gamma_{\rm 1s}$ (e.g., ZK04).

Now, the compressive ratio can be derived directly from equation
(\ref{jump1}), i.e.

\be
\frac{n_2}{n_1} = \frac{u_{\rm 1s}(\gamma_{21},\sigma)}
{u_{\rm 2s}(\gamma_{21},\sigma)} = \gamma_{21} +\frac{[u_{\rm
2s}^2(\gamma_{21},\sigma)+1]^{1/2}} {u_{\rm
2s}(\gamma_{21},\sigma)} (\gamma_{21}^2-1)^{1/2}~. \label{n2/n1}
\ee

In the downstream region, the total pressure includes the
contribution from the thermal pressure $p_2=(\hat\Gamma-1) e_2$
and the magnetic pressure $p_{\rm b,2}=B_{\rm
2s}^2/8\pi\gamma_{\rm 2s}^2$. The ratio between the magnetic
pressure to the thermal one is:
\begin{eqnarray}
\frac{p_{\rm b,2}}{p_2}&=&k^2\left(\frac{\beta_{\rm
1s}}{\beta_{\rm 2s}}\right)^2 \left(\frac{4\pi n_1 m_{\rm p} c^2
\gamma_{\rm 1s}^2 \sigma} {8\pi (\hat\Gamma-1) e_2\gamma_{\rm 2s
}^2}\right)\nonumber\\ &=& {k^2\sigma \over
2(\hat{\Gamma}-1)}{u_{\rm 1s}\over u_{\rm 2s}}({e_2\over n_2m_{\rm
p}c^2})^{-1}, \label{Rp}
\end{eqnarray}
where equations (\ref{EB}) and (\ref{sigma}) have been
used.

For simplicity, in $\S{2}$, we take $\hat{\Gamma}=4/3$. For
$\gamma_{21}\gg1$, $\gamma_{\rm 1s}\approx \gamma_{\rm
2s}\gamma_{21}(1+\beta_{\rm 2s})>\gamma_{\rm 21}\gg1$. Equation
(\ref{Num1}) can be solved analytically (see also Lyubarsky 2003)
\begin{equation}
\beta_{\rm 2s}={1\over
6}(1+\chi+\sqrt{1+14\chi+\chi^2}),\label{Simp1}
\end{equation}
where $\chi\equiv {k^2\sigma\over 1+\sigma}$. We then have
\begin{equation}
{n_2\over n_1}\approx {\gamma_{21}(1+\beta_{\rm 2s})\over
\beta_{\rm 2s}} =\gamma_{21}{7+\chi+\sqrt{1+14\chi+\chi^2}\over
1+\chi+\sqrt{1+14\chi+\chi^2}},\label{Simp2}
\end{equation}
\begin{eqnarray}
{e_2\over n_2m_{\rm p }c^2} &\approx& {3\gamma_{21}\over
4}(\sigma+1)(1+\beta_{\rm 2s})(1-{\chi\over \beta_{\rm
2s}})\nonumber\\ &=&{\gamma_{21} (1+\sigma)\over 8 }
({7+\chi+\sqrt{1+14\chi+\chi^2}})\nonumber\\
&&[1-{6\chi \over 1+\chi+\sqrt{1+14\chi+\chi^2}}],\label{Simp3}
\end{eqnarray}
\begin{eqnarray}
{p_{\rm b,2}\over p_2} &\approx& {2\chi \over (\beta_{\rm 2s}-\chi)}\nonumber\\
&=& {12\chi\over [1-5\chi+\sqrt{1+14\chi+\chi^2}~]}.\label{Simp4}
\end{eqnarray}
Since ${e_2/n_2m_{\rm p }c^2}>0$, equation (\ref{Simp3}) hints
that $\beta_{\rm 2s}>\chi$. For $\sigma\gg1$, $\gamma_{21}\gg1$
and $k=1$, $\chi\approx 1-{1\over \sigma}$, equations (\ref{Simp1}
- \ref{Simp4}) are reduced to
\begin{equation}
\beta_{\rm 2s}\approx 1-{1\over 2\sigma},~ {n_2\over n_1}\approx
2\gamma_{21}, ~ {e_2\over n_2m_{\rm p}c^2}\approx {3\over
4}\gamma_{21},~ {p_{\rm b,2}\over p_2}\approx
4\sigma,\label{Simp21}
\end{equation}
so that $n_2/(4\gamma_{21}+3)n_1\approx 1/2$, $e_2/(\gamma_{21}-1)
n_2m_{\rm p}c^2\approx 3/4$. All these are well consistent with
the numerical results presented in ZK04 (see also our Figure
\ref{Gam21} (b) and (c)). This hints that strong shocks still
exist in the high $\sigma$ regime, as argued by ZK04. On the other
hand, without magnetic dissipation ($k=1$), the upstream magnetic
energy can not be converted to thermal energy and radiation. Since
$p_{\rm b,2}/p_2\approx 4\sigma \gg1$, in the high-$\sigma$
regime, only a tiny amount of total energy (kinetic plus magnetic)
can be used for electron synchrotron radiation.
The GRB radiation efficiency in the $k=1$, $\sigma \gg 1$ model is
therefore too low to interpret the data, which is typically above
$40\%$ (Lloyd-Ronning \& Zhang 2004).

For $\gamma_{21}\gg1$ and $\chi\ll 1$, equations (\ref{Simp1} -
\ref{Simp4}) are reduced to
\begin{eqnarray}
\beta_{\rm 2s}\approx {1+4\chi \over 3},~~~ {n_2\over n_1}\approx
4\gamma_{21}, ~~~\nonumber\\
 {e_2\over n_2m_{\rm p}c^2}\approx
\gamma_{21}(\sigma+1)(1-2\chi),~~~ {p_{\rm b,2}\over p_2}\approx
{6\chi}.\label{Simp22}
\end{eqnarray}
For $\sigma\ll 1$, equations (\ref{Simp22}) are reduced to
$\beta_{\rm 2s}\approx 1/3$, $n_2/n_1\approx 4\gamma_{21}$,
$e_2/n_2m_{\rm p}c^2\approx \gamma_{\rm 21}$ and $p_{\rm
b,2}/p_2\approx 6\sigma\sim 0$. All these are consistent with the
familiar results presented in Blandford \& McKee (1976). As shown
in equations (\ref{Simp21}), for  $k=1$, $\gamma_{21}\gg1$ and
$\sigma\gg1$, $p_{\rm b,2}/p_2\approx 4\sigma$. So, roughly
speaking, for $k=1$ and $\gamma_{21}\gg1$, $p_{\rm b,2}/p_2$ is
linearly proportional to $\sigma$, which is the case (see the thin
dotted line shown in Figure \ref{Sig}(d)).

When writing down eqs.(\ref{jump1}-\ref{jump3}), one has already
assumed that a pair of shocks form as the collision happens. In
order to have a shock, the relative velocity of shells (the
corresponding Lorentz factor is $\gamma_{41}$ following the
standard convention for shock jump condition analysis) should be
faster than the Alfven velocity ($v_{\rm A}=c \sqrt{\sigma \over
1+\sigma}$, the corresponding Lorentz factor $\gamma_A=
\sqrt{1+\sigma}$). So the condition is $\gamma_{41}>\gamma_{\rm
A}$. If this condition is not satisfied, there will be no shock
wave and the two shells would simply bounce off of each other
elastically.

If a shock form, in the rest frame of the upstream the velocity of
the shock front should be faster than $v_{\rm A}$, i.e.,
$\gamma_{\rm 1s}=\gamma_{\rm 2s}\gamma_{21}(1+\beta_{\rm
2s}\beta_{21})>\gamma_{\rm A}$. In the ideal MHD limit, this is
always the case. However, in the presence of magnetic dissipation,
$\gamma_{\rm 1s}>\gamma_{\rm A}$ is not satisfied for an arbitrary
$k$ ($0\leq k\leq 1$) value. This in turn imposes a strict
constraint on the possible choices of $k$. This can be understood
as follows. In the presence of magnetic dissipation, especially
for $k\ll 1$, $\beta_{\rm 2s}\sim 1/3$ or even smaller,
$\gamma_{\rm 1s}>\gamma_{\rm A}$ yields
$\gamma_{21}>{\sqrt{1+\sigma}\over 1.4}$, bearing in mind that
$\gamma_{21}$ is always smaller than $\gamma_{\rm 41}$. Therefore
for a given small value of $k$, $\gamma_{\rm 21}$ has to be larger
than a certain value in order to have a physical solution of the
shock. In other words, a certain value of $\gamma_{21}$ and
$\sigma$, $k$ is constrained within a certain range. When $k=1$
(no dissipation), the shock can happen in the whole
$\gamma_{21}-\sigma$ plane as long as $\gamma_{41} >
\sqrt{1+\sigma}$ is satisfied.

\subsection{Numerical Investigations}
For general cases with arbitrary $\gamma_{21}$, $\sigma$ and $k$,
equation (\ref{Num1}) have to be solved numerically, and equations
(\ref{e/n}), (\ref{n2/n1}) and (\ref{Rp}) can be calculated
correspondingly.

\begin{figure}
\begin{picture}(-20,270)
\put(0,0){\includegraphics{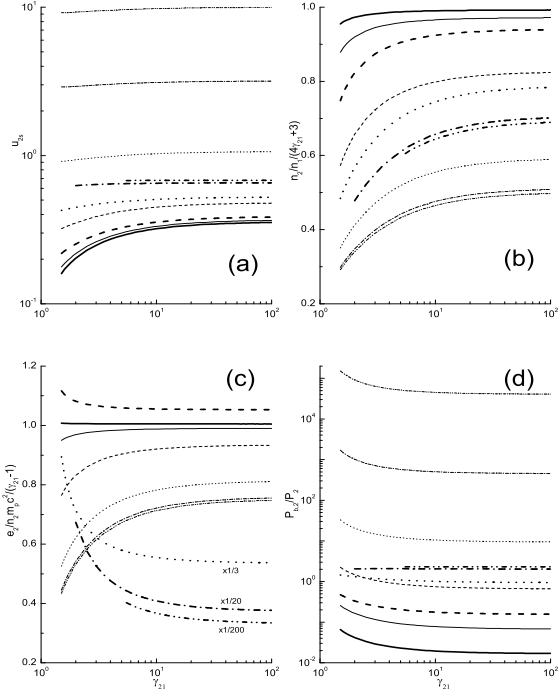}}
\end{picture}
\caption{The variations of four parameters, i.e., $u_{\rm 2s}$,
$e_2/n_2 m_{\rm p} c^2$, $n_2/n_1$ and $p_{\rm b,2}/p_2$, as
functions of $\gamma_{21}$ (sub-figure a, b, c and d,
respectively). The thin lines are for $k=1$, i.e., the ideal MHD
case. The thick lines are for the case with significant magnetic
dissipation, i.e. $k=0.5$. The solid, dashed, dotted, dash-dotted
and dash-dot-dotted lines are for $\sigma=0.01, 0.1, 1, 10, 100$,
respectively. Similar to ZK04, the parameter $e_2 / n_2 m_{\rm p}
c^2$ (thermal Lorentz factor in the shocked, downstream region) is
normalized to $(\gamma_{21}-1)$, and the parameter $n_2/n_1$
(compressive ratio) is normalized to $(4\gamma_{21}+3)$, both
being the values expected in the $\sigma=0$ case. For the
$\sigma=1,~10,~100$ cases, the $e_2 / n_2 m_{\rm p} c^2$ values
are too large to fit into the scale, and we have multiplied the
values by a factor 1/3, 1/20, and 1/200, respectively (see
sub-figure c). For a given $k<1$ and $\sigma$, the condition
$\gamma_{\rm 1s}>\gamma_{\rm A}$ is not satisfied unless
$\gamma_{12}$ is larger than a critical value. So some lines do
not cover the whole horizontal axis scale. Similar situations also
happen in Figure 2.}
 \label{Gam21}
\end{figure}

\begin{figure}
\begin{picture}(-20,280)
\put(0,0){\includegraphics{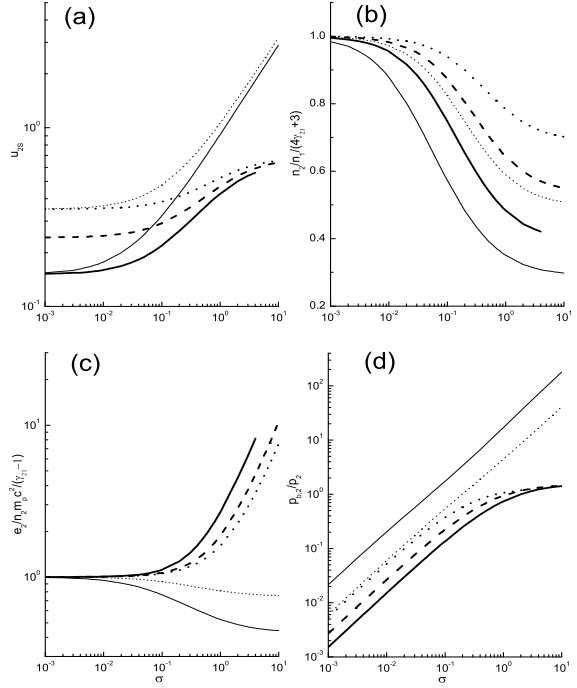}}
\end{picture}
\caption{The variations of four parameters, i.e., $u_{\rm 2s}$,
$e_2/n_2 m_{\rm p} c^2$, $n_2/n_1$ and $p_{\rm b,2}/p_2$, as
functions of $\sigma$ (sub-figure a, b, c and d, respectively).
The thin solid and the thin dotted line represent $k=1$ and
$\gamma_{21}=1.5,~100$, respectively. The thick solid, dashed and
dotted lines are for $k=0.5$ and $\gamma_{21}=1.5, 3.0, 100$,
respectively. Similar to Figure \ref{Gam21}, the parameter $e_2 /
n_2 m_{\rm p} c^2$ is normalized to $(\gamma_{21}-1)$, and the
parameter $n_2/n_1$ is normalized to $(4\gamma_{21}+3)$.}
 \label{Sig}
\end{figure}

Figure \ref{Gam21} and \ref{Sig} show the variations of four
parameters, i.e., $u_{\rm 2s}$, $e_2/n_2 m_{\rm p} c^2$, $n_2/n_1$
and $p_{\rm b,2}/p_2$ as functions of $\gamma_{21}$ (Figure
\ref{Gam21}) and $\sigma$ (Figure \ref{Sig}) respectively. The
thick lines are for $k=0.5$ (the case with significant magnetic
dissipation), and the thin lines are for the $k=1$ (the ideal MHD
limit case). In Figure \ref{Gam21}, the thick solid, dashed,
dotted, dash-dotted and dash-dot-dotted lines are for
$\sigma=0.01, 0.1, 1, 10, 100$, respectively.  The thin solid and
dotted lines are for $\sigma=0.01,100$, respectively. The
parameter $e_2 / n_2 m_{\rm p} c^2$ (thermal Lorentz factor in the
shocked, downstream region) is normalized to $(\gamma_{21}-1)$,
and the parameter $n_2/n_1$ (compressive ratio) is normalized to
$(4\gamma_{21}+3)$, both being the values expected in the
$\sigma=0$ case (Blandford \& McKee 1976). For the
$\sigma=1,~10,~100$ cases, the $e_2 / n_2 m_{\rm p} c^2$ values
are too large to fit into the scale, and we have multiplied the
values by a factor 1/3, 1/20, and 1/200, respectively. In Figure
\ref{Sig}, the thick solid, dashed and dotted lines are for
$k=0.5$, $\gamma_{21}=1.5,~3,~100$, respectively, and the thin
solid and dotted lines are for $k=1$, $\gamma_{21}=1.5,~100$,
respectively. Similar to Figure \ref{Gam21}, the parameter $e_2 /
n_2 m_{\rm p} c^2$ is normalized to $(\gamma_{21}-1)$, and the
parameter $n_2/n_1$ is normalized to $(4\gamma_{21}+3)$.

\begin{figure}
\begin{picture}(-40,210)
\put(0,0){\includegraphics{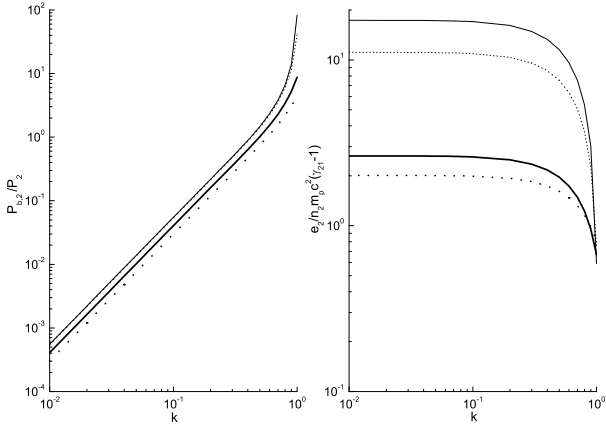}}
\end{picture}
\caption{The variations of two parameters, i.e., $p_{\rm b,2}/p_2$
and $e_2/n_2 m_{\rm p} c^2$, as functions of $k$. The solid line
and dotted line are for $\gamma_{21}=2.5, 100$, respectively. The
thin lines represent $\sigma=10$, the thick lines represent
$\sigma=1$. The parameter $e_2 / n_2 m_{\rm p} c^2$ (thermal
Lorentz factor in the shocked, downstream region) is normalized to
$(\gamma_{21}-1)$, the values expected in the $\sigma=0$ case.}
 \label{Gam22}
\end{figure}

In Figure \ref{Gam21}, when $\gamma_{21}
>10$, nearly all the (normalized) parameters are insensitive to
$\gamma_{21}$ (see also ZK04 for the case of $k=1$), but are
sensitive to $k$ and/or $\sigma$ (see also Figure \ref{Sig} and
Figure \ref{Gam22}). For instance, as shown in Figure
\ref{Gam21}(c), in the absence of magnetic dissipation,
$e_2/(\gamma_{21}-1)n_2m_{\rm p}c^2$ is of order unity for $\sigma
\in [0.01,100]$. But for $k=0.5$, $e_2/(\gamma_{21}-1)n_2m_{\rm
p}c^2\propto (\sigma+1)$ increases linearly with $\sigma$ (see
Figure \ref{Sig}(c)). This is because at higher $\sigma$ values,
more and more magnetic energy is converted to thermal heat while
the total number of leptons is decreasing with $\sigma$, so that
the energy per electron increases rapidly with $\sigma$ (see also
Zhang \& M\'esz\'aros 2002). For $k=1$, $\sigma\gg1$ and
$\gamma_{21}\gg1$, $u_{\rm 2s}\simeq \sqrt{\sigma}$. But for
$k=0.5$, the increase of $u_{\rm 2s}$ with $\sigma$ is only
slightly (see Figure \ref{Gam21}(a), \ref{Sig}(a) for detail). A
similar situation is evident in Figure \ref{Gam21}(d) and
\ref{Sig}(d). For $k=1$, $p_{\rm b,2}/p_2\approx 4\sigma$
increases linearly with $\sigma$. However, for $k=0.5$, the
resulting $p_{\rm b,2}/p_2$ is always of order unity for
$\sigma\gg1$, which hints that significant part of upstream
magnetic energy can be converted into the downstream thermal
energy, no matter how large the $\sigma$ is.

To see the impact of magnetic dissipation on the dynamic evolution
of the shock more clearly, in Figure \ref{Gam22} we plot the
variables $e_2/n_2 m_{\rm p} c^2$ and $p_{\rm b,2}/p_2$ as
functions of $k$. The solid and the dotted lines are for
$\gamma_{21}=2.5, 100$, respectively. The thin lines represent
$\sigma=10$ (For $\sigma=10$, $\sigma/(1+\sigma)\approx 1$, so the
thin solid line and dotted line nearly overlap), the thick lines
represent $\sigma=1$. The parameter $e_2 / n_2 m_{\rm p} c^2$ has
been normalized to $(\gamma_{21}-1)$, the values expected in the
$\sigma=0$ case. The results presented in Figure \ref{Gam22} are
consistent with our equations (\ref{Simp22}), e.g., for $k<0.5$
and $\sigma\gg1$, $e_2/n_2m_{\rm p}c^2\approx
(1-2\chi)(\sigma+1)\gamma_{21}\gg (\gamma_{21}-1)$ and $p_{\rm
b,2}/p_2\approx 6\chi\propto k^2$.

\begin{figure}
\begin{picture}(-20,220)
\put(0,0){\includegraphics{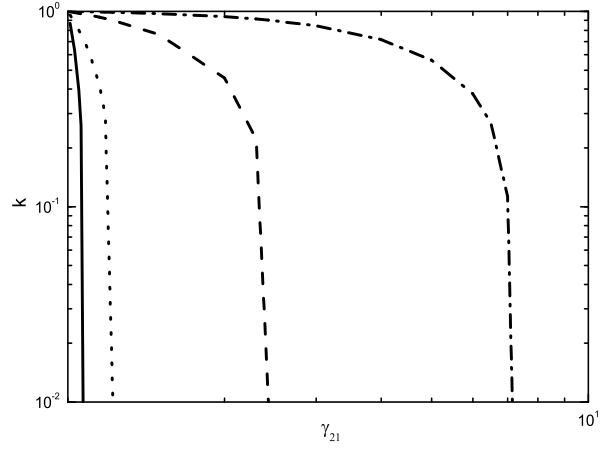}}
\end{picture}
\caption{The condition $\gamma_{\rm 1s}>\gamma_{\rm A}$ poses a
constraint on the range of $k$ as a function of $\gamma_{21}$ and
$\sigma$. The solid, dotted, dashed and dash-dotted lines are for
$\sigma=0.3,~1,~10,~100$ respectively. Above each line is the
physically allowed region.}
 \label{SF}
\end{figure}

As mentioned before, the condition $\gamma_{\rm 1s}>\gamma_{\rm
A}$ imposes a strict constraint on $k$ for a given value of
$\gamma_{21}$ and $\sigma$. Here we discuss it in more detail. The
condition $\gamma_{1s}=\gamma_{\rm 2s}\gamma_{21}(1+\beta_{\rm
2s}\beta_{21})>\gamma_{\rm A}$ is plotted in Figure \ref{SF},
where the solid, dotted, dashed and dash-dotted lines are for
$\sigma=0.3,~1,~10,~100$ respectively. Above each line defined by
a particular $\sigma$ and an arbitrary $k$, a dissipative shock is
allowed. Given a particular $\sigma$ value (i.e. for a particular
line), one can constrain the $k$ value range given a $\gamma_{21}$
value. For a small enough $\gamma_{21}$, each $\gamma_{21}$ value
corresponds to a minimum $k$ value, so that the shock can not be
too dissipative. As $\sigma$ decreases, the restriction is
progressively weaker.

\section{Internal shocks powered by the collision between two
magnetized shells}

\subsection{The internal shocks}
We now use the results obtained in \S2 to study internal shocks.
Consider a faster, trailing, ultra-relativistic, magnetized shell
($\Gamma_{\rm f}$) hits a slower, leading, magnetized shell
($\Gamma_{\rm s}$), where $\Gamma_{\rm f}$ and $\Gamma_{\rm s}$
represent the Lorentz factor of the fast and slow shells (measured
by the observer), respectively. The corresponding velocities for
the two shells are $\beta_{\rm f}$ and $\beta_{\rm s}$,
respectively. Upon collision, two shocks form, i.e. a reverse
shock (RS) propagating into the fast shell and a forward shock
(FS) expanding into the slow shell. The shocks increase the
thermal and magnetic densities of both shells. There are four
regions in this system, i.e. the unshocked slow shell (1), the
shocked slow shell (2), the shocked fast shell (3), the unshocked
fast shell (4). A contact discontinuity separates the shocked fast
shell material and the shocked slow shell material. In the
following analysis, velocities $\beta_{\Gamma_{\rm i}}$ and their
corresponding Lorentz factors $\Gamma_{\rm
i}=(1-\beta_{\Gamma_{\rm i}}^2)^{-1/2}$ ($\Gamma_1\equiv
\Gamma_{\rm s}$ and $\Gamma_4\equiv \Gamma_{\rm f}$) are measured
in the observer frame. Thermodynamic quantities, e.g., $n_{\rm
i}$, $p_{\rm i}$, $e_{\rm i}$ (particle number density, thermal
pressure, thermal energy density) are measured in the rest frame
of the fluid, so are the magnetic pressure and magnetic energy
density, i.e. $p_{\rm B,i}$ and $e_{\rm B,i}$. In this work, we
assume that these two shells are cold, i.e., the specific enthalpy
$\mu_1=\mu_4=m_{\rm p}c^2$ (see equation (\ref{mu}) for the
definition).

The equation that governs the FS takes the form (with equation
(\ref{Num1}))
\begin{eqnarray}
&&\gamma_{\rm 1}u_{\rm 1}(1+{1\over \sigma_1})[\beta_{\rm
1}-\beta_{\rm 2}-{(\hat{\Gamma}_2-1)\over \hat{\Gamma}_2}{1\over
u_{\rm 2}\gamma_{\rm 2}}]+{1\over
2}\nonumber\\
&&=k_2^2{(2-\hat{\Gamma}_2)\over 2\hat{\Gamma}_2}{u_{\rm 1}^2\over
u_{\rm 2}^2}-{(\hat{\Gamma}_2-1)\over
\hat{\Gamma}_2\sigma_1}{u_{\rm 1}\over u_{\rm 2}},\label{Sol1}
\end{eqnarray}
where $\sigma_1\equiv {B_1^2/[4\pi \gamma_1^2n_1m_{\rm p}c^2]}$,
$B_1$ ($B_2$) denote the FS frame magnetic field strength of
region 1 (2);  $\hat{\Gamma}_2$ is the adiabatic index of the
region 2;  $k_2\equiv {\beta_2 B_2\over \beta_1 B_1}$.

Similarly, for the RS we have
\begin{eqnarray}
&&\gamma_{\rm 4}u_{\rm 4}(1+{1\over \sigma_4})[\beta_{\rm
4}-\beta_{\rm 3}-{(\hat{\Gamma}_3-1)\over \hat{\Gamma}_3}{1\over
u_{\rm 3}\gamma_{\rm 3}}]+{1\over
2}\nonumber\\
&&=k_3^2{(2-\hat{\Gamma}_3)\over 2\hat{\Gamma}_3}{u_{\rm 4}^2\over
u_{\rm 3}^2}-{(\hat{\Gamma}_3-1)\over
\hat{\Gamma}_3\sigma_4}{u_{\rm 4}\over u_{\rm 3}},\label{Sol2}
\end{eqnarray}
where $\sigma_4\equiv {B_4^2/[4\pi\gamma_4^2n_4m_{\rm p}c^2]}$,
$B_3$ ($B_4$) denote the RS frame magnetic field strength of
region 3 (4); $\hat{\Gamma}_3$ is the adiabatic index of the
region 3, $k_3\equiv {\beta_3 B_3\over \beta_4 B_4}$. Here
$\gamma_1$ and $\gamma_2$ ($\gamma_3$ and $\gamma_4$) are the
forward (reverse) shock frame Lorentz factor of the fluids in the
region 1 and 2 (3 and 4), respectively, and $u_{\rm
j}^2=\gamma_{\rm j}^2-1$ ($j=1-4$) are the 4-velocities of the
fluids, $\beta_{\rm j}=u_{\rm j}/\gamma_{\rm j}$\footnote{Notice
that the notations here are different from those in \S2 in that we
have dropped out the subscript ``s'' here for simplicity.}. The
$\gamma_{\rm j}$'s can be parameterized by the $\Gamma_{\rm j}$'s
as follows, i.e., for $\Gamma_{\rm fs}, \Gamma_{\rm rs}\gg1$ and
$\Gamma_{\rm j}\gg1$, one has $2\gamma_1\approx (\Gamma_{\rm
fs}/\Gamma_{\rm s}+\Gamma_{\rm s}/\Gamma_{\rm fs})$, $2\gamma_{\rm
2}\approx (\Gamma_{\rm fs}/\Gamma_{\rm 2}+\Gamma_{\rm
2}/\Gamma_{\rm fs})$, $2\gamma_{\rm 3}\approx (\Gamma_{\rm
rs}/\Gamma_{\rm 3}+\Gamma_{\rm 3}/\Gamma_{\rm rs})$ and
$2\gamma_{\rm 4}\approx (\Gamma_{\rm rs}/\Gamma_{\rm
f}+\Gamma_{\rm f}/\Gamma_{\rm rs})$. Here $\Gamma_{\rm fs}$ and
$\Gamma_{\rm rs}$ are the Lorentz factors of the FS and RS
measured in the observer frame, respectively. These equations in
turn suggest that $\Gamma_{\rm 3}\approx
(\gamma_4-u_4)(\gamma_3+u_3)\Gamma_{\rm f}$ and $\Gamma_{\rm
2}\approx (\gamma_2-u_2)(\gamma_1+u_1)\Gamma_{\rm s}$. Therefore
the equality of the velocities along the contact discontinuity
($\Gamma_2=\Gamma_3$) yields
\begin{equation}
\Gamma_{\rm f}\approx
(\gamma_4+u_4)(\gamma_3-u_3)(\gamma_2-u_2)(\gamma_1+u_1)\Gamma_{\rm
s}.\label{Sol3}
\end{equation}
The equality of the total pressure (the sum of the thermal pressure
and the magnetic pressure) along the contact discontinuity
($P_{\rm 2,tot}=P_{\rm 3,tot}$) yields
\begin{eqnarray}
&\{{\hat{\Gamma}_2-1\over \hat{\Gamma}_2}[{\gamma_1 \over
\gamma_2}(1+\sigma_1(1-k_2^2{\beta_1\over \beta_2}))-{m_{\rm
p}c^2\over \mu_1}]+{k_2^2\sigma_1 u_1 \over 2u_2}\}n_2 \mu_1=\nonumber\\
&\{{\hat{\Gamma}_3-1\over \hat{\Gamma}_3}[{\gamma_4 \over
\gamma_3}(1+\sigma_4(1-k_3^2{\beta_4\over \beta_3}))-{m_{\rm
p}c^2\over \mu_4}]+{k_3^2\sigma_4 u_4 \over 2u_3}\}n_3 \mu_4.
\label{Sol4}
\end{eqnarray}
Equations (\ref{Sol1}-\ref{Sol4}) are the basic formulae for the
following calculations.

In this work, it has been assumed that two shocks form in a
collision between two magnetized ``shells''. The condition for
this to happen is that the relative velocity of the two shells
exceeds the $v_{\rm A}$. If the relative Lorentz factor between
the two shells is $\gamma_{\rm 41}$, the condition can be written
as $\gamma_{\rm 41}>\sqrt{1+\sigma}$.

\subsection{Numerical Results}
\begin{figure}
\begin{picture}(20,240)
\put(0,0){\includegraphics{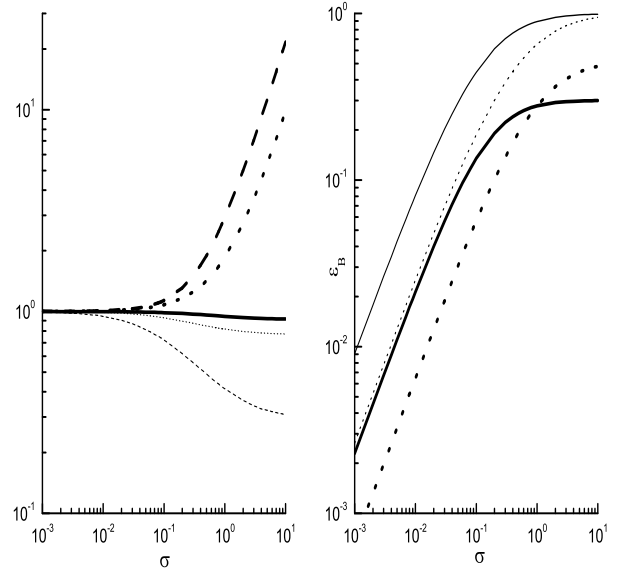}}
\end{picture}
\caption {The bulk and thermal Lorentz factors and the magnetic
equipartition parameter as functions of $\sigma$. $\Gamma_{\rm
f}/\Gamma_{\rm s}=20$ is adopted throughout. The thick lines
represent $k=0.5$, the thin lines represent $k=1$, i.e., the ideal
MHD limit. (a) The solid lines represent the bulk Lorentz factor
of the shocked region, i.e. $\Gamma_2$ ($=\Gamma_3$), normalized
to the value for $\sigma=0$. For $k=0.5$ and 1, the resulting
$\Gamma_2$'s are similar, so that the thin/thick solid lines
overlap. The dotted/dashed lines represent the thermal Lorentz
factors of the shocked baryons in FS/RS regions, respectively,
again normalized to the values for $\sigma=0$; (b)
$\varepsilon_{\rm B}$, the downstream magnetic energy density
normalized to the sum of the downstream thermal energy density and
the magnetic one. The solid/dotted lines correspond to the
$\varepsilon_{\rm B}$ obtained in the FS/RS regions,
respectively.}
 \label{Rad1}
\end{figure}

The problem is complicated and numerical calculations are needed.
For simplicity, we assume that the magnetization parameters
$\sigma$, the dissipation parameter $k$, the rest masses, and the
widths of the shells (measured by the observer) are the same for
both shells, so that $\sigma_1=\sigma_4\equiv\sigma$,
$k_2=k_3\equiv k$, $P_1=P_4=0$ and $f\equiv n_4/n_1=\Gamma_{\rm
s}/\Gamma_{\rm f}<1$ (which implies that the RS is stronger than
the FS). As a numerical example, we assume $\sigma$ ranges from
$10^{-3}$ to $10$, and take $\Gamma_{\rm s}=50$, $\Gamma_{\rm
f}/\Gamma_{\rm s}=1/f=20$, i.e., $\gamma_{41}\approx 10$. We
extend the $\sigma$ range up to 10 in the following calculation,
so that the shock form condition $\gamma_{41} > \sqrt{1+\sigma}$
is always satisfied. Our calculations therefore satisfy the shock
form condition. The dissipation parameter $k$ is taken for two
indicative values, i.e. 1 and 0.5. Another important parameter
involved is the adiabatic index of the shocked material. Here we
treat it self-consistently: we define the thermal Lorentz factor
of the downstream baryons as $\gamma_{\rm th,i'}\equiv e_{\rm
i'}/n_{\rm i'}m_{\rm p}c^2$ ($i'$=2, 3). For $\gamma_{\rm
th,i'}>1$ we take $\hat{\Gamma}_{\rm i'}=4/3$ since both electrons
and protons are relativistic. Otherwise, protons are only
sub-relativistic although electrons are relativistic. In such a
case, we have $\hat{\Gamma}_{\rm i'}=13/9$. The equations
(\ref{Sol1}-\ref{Sol4}) are then solved numerically. With the
resulting $\gamma_{\rm j}~(\rm j=1-4)$, we can calculate
$\Gamma_{\rm 2}$, $\gamma_{\rm th,i'}$ as a function of $\sigma$
and $k$ directly. The numerical results are shown in Figure
\ref{Rad1}(a). Since we are mainly interested in the novel
features introduced by the magnetic fields, we normalize the
relevant parameters with their corresponding value for $\sigma=0$.
For $k=1$ and $\sigma \sim 10^{-3}-10$, the results change only
slightly with respect to those for $\sigma=0$, which implies that
strong internal shocks still exist in the high $\sigma$ case, as
ZK04 argued recently (see the thin lines plotted in Figure
\ref{Rad1}(a) for detail). For $k=0.5$, the results are quite
different. For example, $\gamma_{\rm th,i'}\propto \sigma$ when
$k=0.5$, which can be very high if $\sigma\gg1$ (see the thick
lines plotted in Figure \ref{Rad1}(a) for detail). The main reason
is that for a constant $k$, the total dissipated energy
essentially remains the same for different $\sigma$ values while
the number of leptons decreases sharply as $\propto \sigma^{-1}$,
so that the energy per lepton increases as $\propto \sigma$.
\begin{figure}
\begin{picture}(-20,240)
\put(0,0){\includegraphics{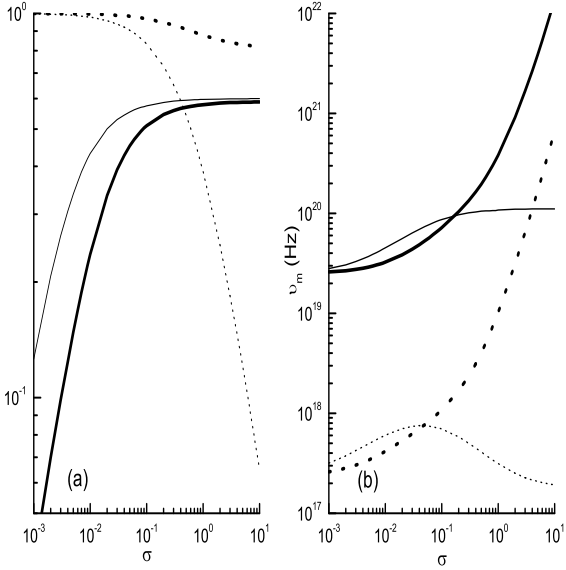}}
\end{picture}
\caption {The net linear polarization degree, the radiation
efficiency, and the typical synchrotron frequency as functions of
$\sigma$. The thick lines represent $k=0.5$, and the thin lines
represent $k=1$, i.e., the ideal MHD limit. $\Gamma_{\rm
f}/\Gamma_{\rm s}=20$ is adopted throughout. (a) The solid line
represents the net linear polarization of the emission from region
3, as a function of $\sigma$ (where $\varepsilon_{\rm B,0}=0.01$
is adopted); The dotted line represents the efficiency of
converting the upstream total energy into the prompt synchrotron
emission frequency as a function of $\sigma$, as normalized to the
value for $\sigma=0$. (b) The observed typical synchrotron
radiation of the shocked electrons in region 2 (the dotted line)
and 3 (the solid line). Other parameters adopted here are
$\varepsilon_{\rm e}=0.5$, $R=10^{13}{\rm cm}$, $\varepsilon_{\rm
B,0}=0.01$, $p=2.5$, and $L_0=10^{52}{\rm ergs}$. The cosmological
redshift correction effect is not taken into account.}
 \label{Rad2}
\end{figure}

The downstream magnetic field is amplified effectively. For
convenience, we define the ratio of magnetic energy density to the
sum of the thermal energy density and the magnetic one as
\begin{eqnarray}
\varepsilon_{\rm B,i'} &\equiv & {e_{\rm B,i'}\over e_{\rm
B,i'}+e_{\rm i'}}\nonumber\\
&=&{{\sigma k^2u_{\rm i''}\over
2u_{\rm i'}}\over {\sigma k^2u_{\rm i''}\over 2u_{\rm i'}}+{1\over
\hat{\Gamma}_{\rm i'}}[{\gamma_{\rm i''}\over \gamma_{\rm
i'}}(1+\sigma (1-k^2{\beta_1\over \beta_2}))-1]}.\label{Epsb}
\end{eqnarray}
Please note that unless $\sigma\ll 1$, it is much different from
the familiar notation $\varepsilon_{\rm B,0}$, the fraction of the
shock-generated magnetic energy density to the total
shock-dissipated energy density in the non-magnetized fireball
model. Here $i''=1,~4$, which corresponds to $i'=2,3$,
respectively. The numerical results are shown in Figure
\ref{Rad1}(b). For $k=1$, the magnetic energy is not dissipated,
so the downstream magnetic energy is strong. For instance, for
$\sigma\geq 3\times 10^{-3}$, the downstream $\epsilon_{\rm
B,i'}\geq 10^{-2}$, which is strong enough to match what is needed
in the internal shock synchrotron model of GRBs (Guetta et al.
2001). For $\sigma\gg1$, $\varepsilon_{\rm B,i'}$ reaches a
asymptotic value $\simeq 1.0$, since the downstream magnetic
energy density dominates the total ones. However, for $k=0.5$,
i.e., significant part of magnetic energy has been dissipated, the
resulting magnetic energy is much weaker than that of $k=1$. For
$\sigma\gg1$, the resulting $\varepsilon_{\rm B,i'}$ reaches a
asymptotic value $\simeq 0.4$.

If our line of sight to the ejecta is not very near the edge of
the cone, due to the beaming effect, the viewed outflow is
essentially axis-symmetric. As a result, the net polarization
contributed by the random magnetic field is nearly zero. On the
other hand, the existence of the ordered magnetic field component
in the ejecta likely results in a net linear polarization (e.g.
Granot 2003). The observed net linear polarization can be
expressed as (Granot \& K\"{o}nigl 2003; Fan et al. 2004)
\begin{equation}
\Pi_{\rm net}\approx 0.60{b^2\over b^2+1},
\end{equation}
where $b^2$ is the ratio of the ordered magnetic energy density to
the random one. In the current case, for the region 3 (Relatively
speaking, the radiation comes from the region 2 is weaker and
softer, which contributes little to the $\gamma-$ray emission. So
we mainly focus on the region 3), $b^2=\varepsilon_{\rm
B,3}/(1-\varepsilon_{\rm B,3})\varepsilon_{\rm B,0}$. We take the
typical value of the random field equipartition parameter as
$\varepsilon_{\rm B,0}\sim 10^{-2}$ (e.g. Guetta et al. 2001). The
net linear polarization degree is calculated in Figure
\ref{Rad2}(a) (the solid lines). We can see that only
$\sigma>3.0\times10^{-3}$ is required to produce $\Pi_{\rm
net}>30\%$ for the $k=1$ case, while for $k=0.5$ one requires
$\sigma>0.02$. This more demanding requirement is simply due to
that much of the downstream magnetic energy has been dissipated
for $k=0.5$.

Assuming that the shocked electrons in the region $i'$ have a
power law distribution in energy, i.e. $dn/d\gamma_{\rm
e,i'}\propto \gamma_{\rm e,i'}^{\rm -p}~(\gamma_{\rm
e,i'}>\gamma_{\rm e,i',m})$ with $p>2$, we have $\gamma_{\rm
e,i',m}\approx \varepsilon_{\rm e}\gamma_{\rm th,i'}[(p-2)/(p-1)]
(m_{\rm p}/m_{\rm e})$. As usual, we assume that the internal
shocks take place at a radius $R\sim 10^{13}{\rm cm}$ and the wind
luminosity is $L_0=10^{52}{\rm ergs~s^{-1}}$. We can then
calculate the typical synchrotron radiation frequency $\nu_{\rm
m,i'}={eB_{\rm i'}\gamma_{\rm e,i',m}^2\Gamma_2/2\pi m_{\rm e}c}$.
Both the ordered magnetic field component and the random magnetic
component are taken into account.

Another important parameter involved is the conversion efficiency
of the internal shock, i.e., the fraction of the total upstream
energy (kinetic plus magnetic) that is converted into the
downstream thermal energy. This quantity is directly related to
the GRB radiation efficiency through the parameter $\epsilon_e$,
i.e. the energy which is transported to electrons can be radiated
effectively in the fast-cooling regime (which is justified in the
GRB prompt emission phase). Since the emission of the region 3 is
much stronger and harder than that of the region 2 (see
\ref{Rad2}(b)), here we mainly consider the former component. The
downstream thermal energy per baryon (in the observer frame) is
$e_{\rm th,3}\approx \gamma_{\rm th,3}\Gamma_2m_{\rm p}c^2$, and
the total number of baryons involved is $N_b=L_0\delta
t/\Gamma_{\rm f}(1+\sigma)m_{\rm p}c^2$, where $E_{\rm
tot}=L_0\delta t$ is the total upstream energy (in the observer
frame), and the factor $1/(1+\sigma)$ represents the fraction of
the kinetic energy to the total energy. The conversion efficiency
can be then written as $\epsilon=e_{\rm th,3} N_b/E_{\rm
tot}\epsilon_{\rm e}\gamma_{\rm th,3}\Gamma_2/\Gamma_{\rm
f}(1+\sigma)$.
This efficiency is plotted against $\sigma$ in Figure
\ref{Rad2}(a) (dotted lines) for both $k=1$ (thin dotted) and
$k=0.5$ (thick dotted). Since we are mainly interested in the
novel features introduced by the magnetic fields, the efficiency
has been normalized to the corresponding value for $\sigma=0$. We
can see that $\epsilon$ generally decreases with $\sigma$. For
$k=1$, the decrease is drastic so that the radiative efficiency
for $\sigma \gg 1$ is too low to interpret the GRB data
(Panaitescu \& Kumar 2001; Lloyd-Ronning \& Zhang 2004). For
$k=0.5$, one can still retain a high efficiency in the
high-$\sigma$ regime.

In Figure \ref{Rad2}(b) we plot the typical synchrotron frequency
as a function of $\sigma$ for both the RS and the FS (each
calculated for both $k=1$ and $k=0.5$). We find that for the
typical parameters adopted here, the forward shock emission has
both a lower frequency and a lower luminosity than the RS. For
$k=1$, the RS emission frequency is about $2\times 10^{20}{\rm
Hz}$, which is insensitive to $\sigma$ and matches the current
observation. However, as discussed above, the $\sigma\geq 1$
regime is disfavored due to the low efficiency involved. For
$k=0.5$, the downstream magnetic energy can be converted into the
prompt $\gamma-$ray emission effectively, but for $\sigma\gg1$,
the resulting RS emission frequency $\nu_{\rm m}\propto \sigma^2$
is much harder than the BATSE band (Fig.\ref{Rad2}(b); see also
Zhang \& M\'{e}sz\'{a}ros 2002).

\section{Discussions}

In the standard fireball model of GRBs, the prompt $\gamma-$ray
emission is believed to be powered by internal shocks (e.g.,
M\'{e}sz\'{a}ros 2002). That model is successful to interpret the
GRB variability (e.g., Kobayashi et al. 1997) and some empirical
relations (e.g., Zhang \& M\'{e}sz\'{a}ros 2002). However, if the
magnetic field involved in the internal shock region is random
(e.g. generated within the shocked region due to plasma
instability), it is difficult to account for the observed high
linear polarization of GRB021206 (Waxman 2003; Granot 2003). If
the involved shells are magnetized, even if the magnetization
parameter is only mild (e.g. $\sigma > 10^{-3}$ without magnetic
dissipation and $\sigma > 0.02$ for substantial magnetic
dissipation), the amplified downstream ordered magnetic field is
strong enough to dominate the random component and generate a
significant degree of linear polarization (see Figure
\ref{Rad2}(a)). For a lower magnetization parameter, e.g.,
$\sigma\ll 10^{-3}$, no observable net high linear polarization is
produced unless some other geometry effects are taken into
account.

Through introducing a ``magnetic dissipation parameter'', $k$
(Lyubarsky 2003), the MHD jump condition with magnetization is
solved, and the parameterized expressions and numerical
calculations for the compressive ratio ($n_{\rm d}/n_{\rm u}$),
the magnetic pressure - to - thermal pressure ratio ($p_{\rm
b,d}/p_{\rm d}$), and the downstream mean random Lorentz factor
($e_{\rm d}/n_{\rm d}m_{\rm p}c^2$) (where the subscript $u$, $d$
represent the upstream and downstream respectively) are presented
for various $\gamma_{21}$, $\sigma$ and $k$ values (See \S{2}). In
this work, we treat $\sigma$ and $k$ as independent parameters and
investigate the dependence of the solution on both parameters. As
expected, we found that the results obtained for shocks with
magnetic energy dissipation is very different from the familiar
one obtained in the ideal MHD limit. The introduction of the $k$
parameter manifests our ignorance of the poorly known magnetic
dissipation process, and its value is poorly constrained. In
reality, $k$ may be correlated with $\sigma$ and $\gamma_{21}$.
For example, in the $\sigma \rightarrow 0$ limit one has $k
\rightarrow 1$. Also in Fig. \ref{SF} we have shown that the lower
limit of $k$ is jointly determined by $\gamma_{21}$ and $\sigma$.
However, lacking a theory for magnetic dissipation, we simply
treat $k$ as a free parameter as long as it satisfies the
condition shown in Fig. \ref{SF}. With the general jump
conditions, the GRB internal shocks with moderate magnetization
($10^{-3}<\sigma<10$) have been calculated. Various
considerations/constraints allow us to narrow down the $\sigma$
range for a possible GRB model. We show that for $k=1$, i.e., the
ideal MHD limit, strong internal shocks still exist in the high
$\sigma$ case ($\sigma\gg 1$). However, in the $\sigma \gg 1$
regime, the upstream magnetic energy can not be converted into the
downstream internal energy effectively, resulting in a very low
radiation efficiency inconsistent with the current GRBs data
(Panaitescu \& Kumar 2001; Lloyd-Ronning \& Zhang 2004). For a
significant magnetic dissipation case (e.g. $k\leq 0.5$), the
upstream magnetic energy can be converted into the prompt
$\gamma-$ray emission effectively for an arbitrary $\sigma$, but
for $\sigma\gg1$, the observed frequency ($\propto \sigma^2$) is
much harder than what we observe giving a typical internal shock
radius. We therefore disfavor a $\sigma$ value much greater than
unity. We note that too high a typical radiation frequency is a
common feature for any high-$\sigma$ model, and the problem may be
remedied by considering a possible pair cascade in the magnetic
dissipation region (e.g. Zhang \& M\'esz\'aros 2002; the pair
emission in the internal shocks can be found in Fan \& Wei 2004
and Li \& Song 2004). Developing a detailed pair-dominated model
in the high-$\sigma$ regime is however beyond the scope of the
present work.

At the low-$\sigma$ side, if the claimed high linear polarization
in GRB 021206 is true (Coburn \& Boggs 2003; cf. Rutledge \& Fox
2004), within the synchrotron model, the required magnetization
parameter is $\sigma > 10^{-3}$ for $k=1$ and $\sigma > 0.02$ for
$k=0.5$ in order to give rise to a $\geq 30\%$ linear
polarization. Modelling early afterglows for GRB 990123 and GRB
021211 generally requires a magnetized flow (Fan et al. 2002;
Zhang et al. 2003; Kumar \& Panaitescu 2003), with $\sigma > 0.1$
(ZK04) for these two bursts. Considering more general cases, we
favor a mildly magnetized fireball model ($10^{-2}<\sigma\ll 1$),
first suggested in Rees \& M\'{e}sz\'{a}ros (1994). A traditional
problem of the internal shock model is its low radiation efficiency
(e.g. Panaitescu, Spada \& M\'esz\'aros 1999; Kumar 1999).
In the presence of a toroidal magnetic field, the efficiency of the
internal shock is even lower in the ideal MHD limit. In view
that the magnetic dissipation process (which is naturally expected
for a moderate $\sigma$ value) can help to solve this problem (see
Figure \ref{Rad2}(a), the thick dashed line for detail),
we suggest that a mildly magnetized internal shock model with
moderate magnetic dissipation is a good candidate to explain the
current GRB prompt emission data.

The mild magnetization (e.g. $\sigma \sim (10^{-2} - 1)$)
preferred in this paper is likely a natural outcome of a realistic
central engine. In a GRB event, a rapidly rotating magnetar-type
(either black hole - torus system or neutron star) central engine
with a surface magnetic field $\sim 10^{15}$ G likely launches a
Poynting flux flow with an isotropic luminosity $\sim
10^{50}-10^{51}{\rm ergs~s^{-1}}$. The cataclysmic event also
involves a hot fireball component due to processes such as
neutrino annihilation, with a typical isotropic luminosity of
$\sim 10^{51}-10^{52}~{\rm ergs~s^{-1}}$. The latter energy
component may be orders of magnitude stronger than or is at least
comparable with the former, so that the picture recommended here
is justified.

\section*{Acknowledgments}

We thank T. Lu, Z. G. Dai, Y. F. Huang, X. Y. Wang \& X. F. Wu for
fruitful discussions. We appreciate the anonymous referee for
helpful comments that enable us to improve the paper
significantly. This work is supported by the National Natural
Science Foundation (grants 10225314 and 10233010) and the National
973 Project on Fundamental Researches of China (NKBRSF G19990754).
B.Z. acknowledges NASA NAG5-13286, and NASA Long Term Space
Astrophysics (NNG04GD51G) for supports.

\begin{appendix}
\section{Derivation of equation (10)}
Equations (\ref{jump1}), (\ref{jump2}) and (\ref{jump3}) can be
rearranged as follows (with (\ref{EB}))
\begin{eqnarray}
n_1 u_{\rm 1s}         &=&  n_2 u_{\rm 2s}, \label{A1}\\
\gamma_{\rm 1s} u_{\rm 1s}w_1 &=& \gamma_{\rm 2s} u_{\rm 2s}w_2,
\label{A2}\\
w_1 u_{\rm 1s}^2+(p_1+{B_1^2/ 8\pi})
   &=&  w_2 u_{\rm 2s}^2+(p_2+{B_2^2/ 8\pi}), \label{A3}
\end{eqnarray}
where $w_{\rm i}\equiv n\mu_{\rm i}+B_{\rm i}^2/4\pi=n_{\rm
i}m_{\rm p}c^2+{\hat{\Gamma_{\rm i}}\over \hat{\Gamma}_{\rm i
}-1}p_{\rm i }+B_{\rm i}^2/4\pi$, $B_{\rm i}$ (i=1,2) are measured
in the comoving frame, $\hat{\Gamma}_{\rm i}$ are the adiabatic
index of region i. With the definition of $w_{\rm i}$, we have
\begin{equation}
p_{\rm i}={\hat{\Gamma}_{\rm i}-1\over \hat{\Gamma}_{\rm
i}}(w_{\rm i}-n_{\rm i}m_{\rm p}c^2-B_{\rm i}^2/4\pi).
\end{equation}
Equation (\ref{A2}) then reads
\begin{eqnarray}
w_1(u_{\rm 1s}^2+{\hat{\Gamma}_{\rm 1}-1\over \hat{\Gamma}_{\rm
1}})-{\hat{\Gamma}_1-1\over \hat{\Gamma}_1}n_1m_{\rm
p}c^2+{2-\hat{\Gamma}_1\over 2\hat{\Gamma}_{1}}{B_{1}^2\over
4\pi}\nonumber\\
=w_2(u_{\rm 2s}^2+{\hat{\Gamma}_{\rm 2}-1\over \hat{\Gamma}_{\rm
2}})-{\hat{\Gamma}_2-1\over \hat{\Gamma}_2}n_2m_{\rm
p}c^2+{2-\hat{\Gamma}_2\over 2\hat{\Gamma}_{2}}{B_{2}^2\over
4\pi}.\label{NA3}
\end{eqnarray}
On the other hand, equation (\ref{A3}) yields $w_2={\gamma_{\rm
1s}u_{\rm 1s}w_1/ \gamma_{\rm 2s}u_{\rm 2s}}$. Substituting this
into equation (\ref{NA3}), we have
\begin{eqnarray}
&\gamma_{\rm 1s}u_{\rm 1s}w_1(\beta_{\rm 1s}-\beta_{\rm
2s}+{\hat{\Gamma}_1-1\over \hat{\Gamma}_1}{1\over \gamma_{\rm
1s}u_{\rm 1s}}-{\hat{\Gamma}_2-1\over
\hat{\Gamma}_2}{1\over \gamma_{\rm 2s}u_{\rm 2s}})\nonumber\\
&={\hat{\Gamma}_1-1\over \hat{\Gamma}_1}n_1m_{\rm
p}c^2-{\hat{\Gamma}_2-1\over \hat{\Gamma}_2}n_2m_{\rm
p}c^2+{2-\hat{\Gamma}_2\over 2\hat{\Gamma}_2}{B_2^2\over
4\pi}-{2-\hat{\Gamma}_1\over 2\hat{\Gamma}_1}{B_1^2\over 4\pi}.
\end{eqnarray}
For $p_1\ll n_1m_{\rm p}c^2$, which we are interested in here, one
has $w_1\approx n_1m_{\rm p}c^2(1+\sigma)$. Considering
$n_2/n_1=u_{\rm 1s}/u_{\rm 2s}$ and $[B_{\rm 2}/B_{\rm
1}]^2=k^2[u_{\rm 1s}/u_{\rm 2s}]^2$, we have
\begin{eqnarray}
&\gamma_{\rm 1s}u_{\rm 1s}(1+\sigma)(\beta_{\rm 1s}-\beta_{\rm
2s}+{\hat{\Gamma}_1-1\over \hat{\Gamma}_1}{1\over \gamma_{\rm
1s}u_{\rm 1s}}-{\hat{\Gamma}_2-1\over
\hat{\Gamma}_2}{1\over \gamma_{\rm 2s}u_{\rm 2s}})\nonumber\\
&={\hat{\Gamma}_1-1\over \hat{\Gamma}_1}-{\hat{\Gamma}_2-1\over
\hat{\Gamma}_2}{u_{\rm 1s}\over u_{\rm 2s}}+{2-\hat{\Gamma}_2\over
2\hat{\Gamma}_2} k^2({u_{\rm 1s}\over u_{\rm
2s}})^2\sigma-{2-\hat{\Gamma}_1\over 2\hat{\Gamma}_1}\sigma.
\end{eqnarray}
After some simple algebra, we finally have
\begin{eqnarray}
&\gamma_{\rm 1s}u_{\rm 1s}(1+\sigma)(\beta_{\rm 1s}-\beta_{\rm
2s}-{\hat{\Gamma}_2-1\over
\hat{\Gamma}_2}{1\over \gamma_{\rm 2s}u_{\rm 2s}})+{\sigma\over 2}\nonumber\\
&=-{\hat{\Gamma}_2-1\over \hat{\Gamma}_2}{u_{\rm 1s}\over u_{\rm
2s}}+{2-\hat{\Gamma}_2\over 2\hat{\Gamma}_2} k^2({u_{\rm 1s}\over
u_{\rm 2s}})^2\sigma,
\end{eqnarray}
which is equation (\ref{Num1}) in \S{2}.
\end{appendix}
\end{document}